\documentclass[letter]{aa} 
\usepackage{graphicx}
\usepackage{txfonts}
\begin{document}
   \title{A very massive runaway star from Cygnus OB2\thanks{Based on
observations collected at the Centro Astron\'omico Hispano-Alem\'an
(CAHA) at Calar Alto, operated jointly by the Max-Planck Institut
f\"ur Astronomie and the Instituto de Astrof\'\i sica de Andaluc\'\i
a (CSIC).}}


   \author{F. Comer\'on
          \inst{1}
          \and
          A. Pasquali\inst{2}
          }

   \offprints{F. Comer\'on}

   \institute{ESO, Karl-Schwarzschild-Str. 2, D-85748 Garching bei
   M\"unchen, Germany\\
              \email{fcomeron@eso.org}
         \and
              Max-Planck Institut f\"ur Astronomie, K\"onigstuhl 17,
              D-69117 Heidelberg, Germany\\
            \email{pasquali@mpia.de}
             }

   \date{Received; accepted}

  \abstract
   {}
   {We analyze the available information on the star BD$+43^\circ \ 3654$
   to investigate the possibility that it may have had its origin in the
   massive OB association Cygnus~OB2.}
   {We present new spectroscopic observations allowing a reliable spectral
   classification of the star, and discuss existing MSX observations of its
   associated bow shock and astrometric information not previously studied.}
   {Our observations reveal that BD$+43^\circ \ 3654$ is a very early and
   luminous star of spectral type O4If, with an estimated mass of
   $(70 \pm 15)$~M$_\odot$ and an age
   of $\sim 1.6$~Myr. The high spatial resolution of the MSX observations
   allows us to determine its direction of motion in the plane of the sky
   by means of the symmetry axis of the well-defined bow shock, which
   matches well the orientation expected from the proper motion. Tracing
   back its path across the sky we find that BD$+43^\circ \ 3654$
   was located near the central, densest region of Cygnus~OB2 at a
   time in the past similar to its estimated age.}
   {BD$+43^\circ \ 3654$ turns out to be one of the three most massive
   runaway stars
   known, and it most likely formed in the central region of Cygnus~OB2.
   A runaway formation mechanism by means of dynamical ejection is
   consistent with our results.}


   \maketitle
%

\section{Introduction}

  Current theories for the formation of massive stars stress the
importance of the dense cluster environment in which most of them,
if not all, form (Bonnell et al.~\cite{bonnell07}). Dynamical
interactions at the centers of massive star forming regions lead to
captures forming binary systems, ejections, mass segregation, and
possibly coalescence. A remarkable byproduct of the dynamical
interactions in dense clusters of massive stars is the relatively
large abundance of runaway O-type stars, which amount to almost
$\sim 10$\% of the known O-type stars in the solar vicinity (see
Ma\'\i z-Apell\'aniz~et al.~(\cite{maiz04}) for a recent census).
Runaway stars, characterized by their large spatial velocities, can
form either by dynamical ejection from a dense cluster (Poveda et
al.~\cite{poveda67}, Leonard \& Duncan~\cite{leonard88},
\cite{leonard90}) or by the explosion as supernova of a member of a
close massive binary (Blaauw~\cite{blaauw61}, van Rensbergen et
al.~\cite{vanrensbergen96}, de Donder et al.~\cite{dedonder97}).
Evidence for actual examples resulting from both mechanisms exists
(Hoogerwerf et al.~\cite{hoogerwerf00}, \cite{hoogerwerf01}), and
both are a consequence of the special conditions in which massive
star formation takes place. On the one hand, the high stellar
density of the parental cluster facilitates the dynamical ejection
scenario. On the other hand, the supernova scenario is favored by
the high frequency of binaries with high mass ratios among massive
stars (Garmany et al.~\cite{garmany82}, Preibisch et
al.~\cite{preibisch99}), which may be a consequence of dynamical
capture followed by accretion and orbital evolution (Bate et
al.~\cite{bate03}).

  \object{Cygnus~OB2}, the most massive OB association of the solar
neighbourhood (Kn\"odlseder~\cite{knodlseder00},
\cite{knodlseder03}, Comer\'on et al.~\cite{comeron02}, and
references therein), should be the source of numerous runaway stars
given its rich content in massive stars, which includes the massive
multiple system Cyg~OB2~8 near its center. Unfortunately, few
studies to the date have addressed its possible runaway population,
with the exception of the recent radial velocity survey of Kiminki
et al.~(\cite{kiminki07}) in which no runaway candidate has been
identified until now. Comer\'on et al.~(\cite{comeron94},
\cite{comeron98a}) pointed out the existence of large-scale
kinematical peculiarities in the Cygnus region, most likely related
to the presence of Cygnus~OB2, as shown by {\sl Hipparcos} proper
motions. Although they interpreted their results in terms of
triggered star formation (Elmegreen~\cite{elmegreen98}), at least
some of the stars that they identified as moving away from
Cygnus~OB2 might be actual runaways formed by either of the two
mechanisms listed above.

  In this paper we report the identification of a very high mass
runaway star, \object{BD$+43^\circ \ 3654$}, very probably ejected
from Cygnus~OB2. The star had been already identified as a likely
runaway by van Buren \& McCray~(\cite{vanburen88}) based on the
existence of an apparent bow shock in IRAS images, caused by the
interaction of its stellar wind with the local interstellar medium.
Here we present the first spectroscopic observations of the star,
which show it to be a very early Of-type supergiant. We also present
proper motion data and higher resolution MSX images leading to a
more detailed analysis, which strongly supports an origin at the
core of Cygnus~OB2.

\section{Observations\label{observations}}

  The spectrum presented here was obtained in the course of a
project aimed at producing spectral classifications of previously
unknown, photometrically selected new OB stars in the surroundings
of Cygnus~OB2. The photometry in the $BRJHK_S$ bands was taken from
the Naval Observatory Merged Astrometric Dataset (NOMAD) catalog
(Zacharias et al.~\cite{zacharias04}), which combines astrometry and
photometry from the Hipparcos, Tycho-2, UCAC2, USNO-B1.0, and 2MASS
catalogs. The spectroscopic observations were carried out with the
2.2m telescope at the German-Spanish Astronomical Center on Calar
Alto (Spain) using the CAFOS visible imager and spectrograph. A 1"5
slit combined with the B-100 grism, providing a resolution $\lambda
/ \Delta \lambda = 800$ in the blue part of the visible spectrum,
were used. The exposure time was 900~s. The spectrum was reduced,
extracted, and wavelength calibrated using standard
IRAF\footnote{IRAF is distributed by NOAO, which is operated by the
Association of Universities for Research in Astronomy, Inc., under
contract to the National Science Foundation.} tasks under the
ONEDSPEC package, and it was ratioed by a sixth-degree polynomial
fit to the continuum in order to remove the steep slope due to the
strong extinction towards the star.

\section{Results\label{results}}

\subsection{Stellar classification, properties, and
kinematics\label{properties}}

\begin{figure}
\centering
   \includegraphics[width=8cm]{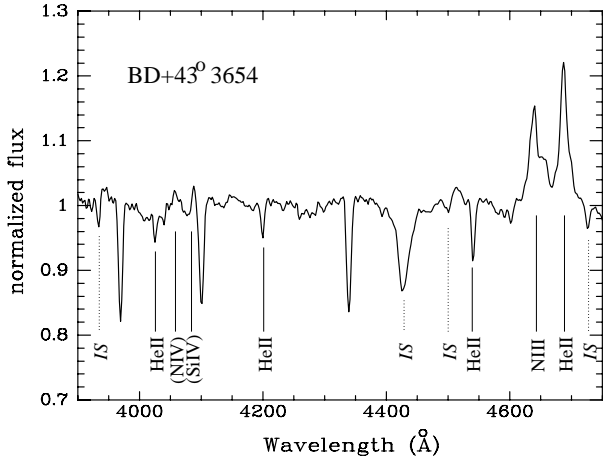}
      \caption{Spectrum of BD$+43^\circ \ 3654$ showing the main
absorption lines used for spectral classification and the prominent
emission of NIII and HeII. The prominent, unlabeled features are
H$\gamma$, H$\delta$, and H$\epsilon$. Interstellar absorption
features are indicated by dotted lines. The locations where one
might expect to detect NIV and SiIV transitions are also indicated.}
         \label{spec}
\end{figure}

  Although the identification of BD$+43^\circ \ 3654$ as a likely
runaway star dates back to van Buren \& McCray~(\cite{vanburen88}),
no spectral classification is available in that work. Subsequent
papers by van Buren et al.~(\cite{vanburen95}) and Noriega-Crespo et
al.~(\cite{noriega97}) refer to the star as a unspecified B-type but
do not report dedicated observations, and no other spectroscopic
classification appears to be available in the literature apart from
a generic classification as 'OB reddened' in the LS catalog (Hardorp
et al.~\cite{hardorp64}). The spectrum presented here is thus the
first one allowing an accurate spectral classification of
BD$+43^\circ \ 3654$ and the estimate of its physical parameters.

  The most obvious spectroscopic feature of BD$+43^\circ \ 3654$ is
the presence of intense emission in the NIII and HeII lines, and
possibly also in NIV and SiIV, clearly indicating that it is a Of
star. HeII lines are also prominent in absorption, and together with
the absence of HeI lines indicates a spectral type earlier than O5.
Absorption bands due to interstellar absorption, CaII and diffuse
interstellar bands, are also strong due to the high extinction
towards the star. Based on comparison with the atlas of Walborn \&
Kirkpatrick~(\cite{walborn90}), we classify the star as O4If. Using
intrinsic colors of early-type stars from
Tokunaga~(\cite{tokunaga00}) and the 2MASS $HK_S$ photometry from
the NOMAD catalog, we estimate a $K$-band extinction $A_K =
0.57$~mag.

  A summary of previous distance determinations to Cygnus~OB2 has
been presented by Hanson~(\cite{hanson03}). Based on her results, we
adopt her favored distance modulus $DM = 10.8$ corresponding to a
distance of 1450~pc, with an estimated uncertainty of $\pm 0.4$
based on the results of previous determinations summarized in that
work. Assuming that BD$+43^\circ \ 3654$ is approximately at the
same distance from the Sun as Cygnus~OB2, we derive its absolute
magnitude as

$$ M_V = K_S - A_K + (V-K_S)_0 - DM = - 6.3 \pm 0.5  \eqno(1)$$

\noindent where the 0.5~mag uncertainty includes as the dominant
source the quoted uncertainty in the distance modulus and the
contribution of error in the derivation of the extinction. We
estimate the latter to be 0.2~mag based on the quality of the fit of
a reddened O4-type spectral energy distribution to the measured
$BRJHK_S$ photometry. The contribution of the errors in the
broad-band photometry is negligible as compared to those other two
sources.

\begin{figure}
\centering
   \includegraphics[width=8cm]{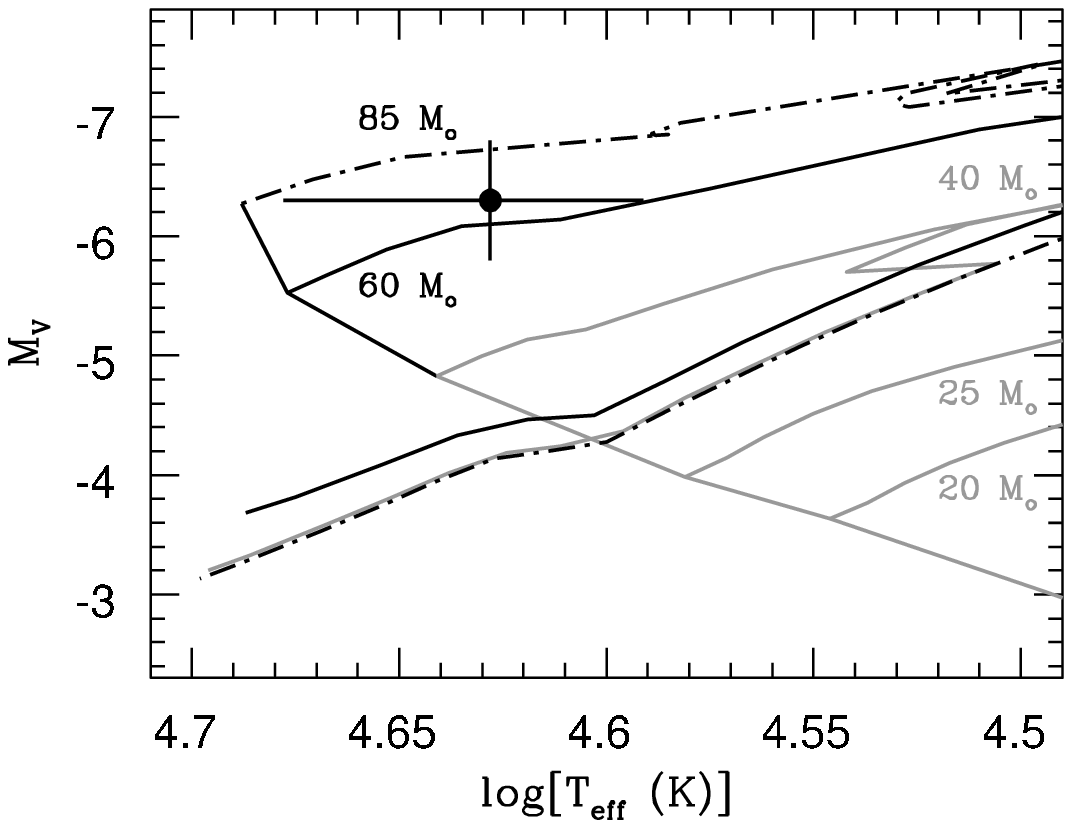}
      \caption{Comparison between the position of BD$+43^\circ \ 3654$
and the evolutionary tracks for very massive stars of Meynet et
al.~(\cite{meynet94}).}
         \label{hrd}
\end{figure}

  Different calibrations of the stellar parameters of O stars can be
found in literature to estimate the mass and the age of BD$+43^\circ
\ 3654$. These calibrations are based on a different treatment of
the stellar model atmospheres, depending on whether non-LTE
conditions, line-blanketing effects and stellar winds are taken into
account. For an O4 supergiant in the Milky Way (i.e. of solar
metallicity), Martins et al.~(\cite{martins05}) provide an effective
temperature $T_{eff}$ = 40702~K; Repolust et al.~(\cite{repolust04})
estimate a colder $T_{eff} = 39000$~K, while Vacca et
al.~(\cite{vacca96}) give $T_{eff} = 47690$~K. We have adopted the
average of those three calibrations, 42464~K, as the temperature for
BD$+43^\circ \ 3654$, considering as the uncertainty the range of
temperatures spanned by those models. This uncertainty is larger
than the temperature difference between the subtypes O4 and O5, and
between types O4I and O4V, for any given calibration (see e.g.
Martins et al.~\cite{martins05}). The same is true for the effects
of metallicity, which are hardly noticeable even when metal
abundances change by a factor of 10. This is clearly shown in
Fig.~15 of Heap et al.~(\cite{heap06}), where the
temperature-spectral type relationships from different calibrations
involving both galactic and Small Magellanic Cloud O stars are
compared. Therefore, plausible uncertainties in either our spectral
classification or in our assumption of solar metallicity for
BD$+43^\circ \ 3654$ do not significantly alter the size of the
error bars in Fig.~\ref{hrd}. The absolute magnitudes $M_V$ obtained
by all three models are very similar, with an average of $M_V =
-6.36$ and individual determinations deviating by less than 0.05~mag
from that value. This is remarkably close to the value that we
obtain from the photometry of BD$+43^\circ \ 3654$ and the
assumption that its distance modulus is the same obtained by
Hanson~(\cite{hanson03}) for Cygnus~OB2, thus supporting our choice
of that distance for the star.

  The position of BD$+43^\circ \ 3654$ on the Herzsprung-Russell (HR)
diagram is shown in Fig.~\ref{hrd}, together with the isochrones
computed by Meynet et al.~(\cite{meynet94}) for solar metallicity
and with enhanced stellar mass loss. These evolutionary tracks are
preferable because they better reproduce the low-luminosity observed
for some Wolf-Rayet stars, the surface chemical composition of WC
and WO stars and the ratio of blue to red supergiants in the star
clusters of the Magellanic Clouds. The isochrones plotted in
Fig.~\ref{hrd} refer to a stellar mass of the progenitor on the main
sequence of $M_i = 20, 25 , 40$~M$_\odot$ (in grey) and $M_i = 60,
85$~M$_\odot$ (in black). The comparison between the observed
properties of BD$+43^\circ \ 3654$ and the isochrones allows us to
estimate an initial mass $M_i \simeq (70 \pm 15)$~M$_\odot$ and an
approximate age of 1.6~Myr. The isochrones do not take into account
stellar rotation, which many studies in the past decade have found
to greatly affect mixing and mass loss, and to be an important
ingredient for stellar evolution (Meynet \& Maeder~\cite{meynet97},
Langer et al.~\cite{langer98}, Heger \& Langer~\cite{heger00},
Meynet \& Maeder~\cite{meynet00}, and references therein). As shown
by Meynet \& Maeder~(\cite{meynet00}), for an initial rotational
velocity of 200-300~km~s$^{-1}$ and solar metallicity isochrones
become brighter by a few tenths of a magnitude and the lifetime in
the H-burning phase increases by 20-30\%. Given the observational
errors on BD$+43^\circ \ 3654$, these changes do no affect
significantly our estimates of the initial mass and age of the star.

  Proper motions for BD$+43^\circ \ 3654$ are available from the
NOMAD catalog, based on measurements by the {\sl Hipparcos}
satellite in the Tycho catalog further refined with previous
ground-based observations. The values listed are $\mu_\alpha \cos
\delta = (-0.4 \pm 0.7)$~mas~yr$^{-1}$ and $\mu_\delta = (+1.3 \pm
1.0)$~mas~yr$^{-1}$. The corresponding values expressed in galactic
coordinates, which are more convenient to derive the spatial
velocity of the star with respect to its local interstellar medium,
are $\mu_l \cos b = (+0.8 \pm 0.9)$~mas~yr$^{-1}$, $\mu_b = (+1.1
\pm 0.8)$~mas~yr$^{-1}$.

  Assuming that the local interestellar medium in the surroundings
of BD$+43^\circ \ 3654$ moves in a circular orbit around the
galactic center, its proper motion $(\mu_l \cos b)_0$, $(\mu_b)_0$
can be described by the first-order approximation to the local
galactic velocity field; see e.g. Scheffler \&
Els\"asser~(\cite{scheffler87}):

$$(\mu_l \cos b)_0 = 0.211 [A \cos 2l \cos b + B \cos b$$
$$+ {U \over D} \sin l - {V \over D} \cos l]  \eqno(2a)$$
$$(\mu_b)_0 = 0.211 [-A \sin 2l \sin b \cos b$$
$$+ {U \over D} \cos l \sin b + {V \over D} \sin l \sin b - {W \over
D} \cos b]  \eqno(2b)$$

\noindent where $A$ and $B$ are the Oort constants in units of
km~s$^{-1}$~kpc$^{-1}$; $U$, $V$, and $W$ are the components of the
solar peculiar motion in the directions toward the galactic center,
the direction of circular galactic rotation, and the North galactic
pole respectively, in km~s$^{-1}$, and $D$ is the distance to the
Sun in kpc. We have adopted $A = -B = 12.5$~km~s$^{-1}$~kpc$^{-1}$,
corresponding to a flat rotation curve with an angular velocity of
25~km~s$^{-1}$~kpc$^{-1}$ and $(U, V, W) = (7, 14, 7)$~km~s$^{-1}$.
The proper motion of BD$+43^\circ \ 3654$ with respect to its local
interstellar medium is then

$$\Delta \mu_l \cos b = \mu_l \cos b - (\mu_l \cos b)_0 =
(5.3 \pm 1.1)~{\rm mas~yr}^{-1}  \eqno(3a)$$
$$\Delta \mu_b = \mu_b - (\mu_b)_0 =
(2.0 \pm 0.9)~{\rm mas~yr}^{-1}  \eqno(3b)$$

\noindent where the uncertainty allows for an error of
2~km~s$^{-1}$~kpc$^{-1}$ in each of $A$, $B$, and 2~km~s$^{-1}$ in
each of $U$, $V$, and $W$. The position angle $\theta$ of the
residual proper motion with respect to the North galactic pole,
counted as positive in the direction of increasing galactic
latitude, is then

$$\theta = \tan^{-1} {{\Delta \mu_l \cos b} \over {\Delta \mu_b}} =
69^\circ 3 \pm 9^\circ 4 \eqno(4)$$

  The component of the spatial velocity on the plane of the sky that we
derive from the residual proper motion at the adopted distance of
1450~pc is $(39.8 \pm 9.8)$~km~s$^{-1}$, which is several times the
sound speed in a warm neutral interstellar medium at a temperature
of $\sim 8000$~K, as expected from the fact that a clear bow shock
is observed ahead of the star in the direction of its motion.

\subsection{The bow shock\label{bow_shock}}

\begin{figure}
\centering
   \includegraphics[width=8cm]{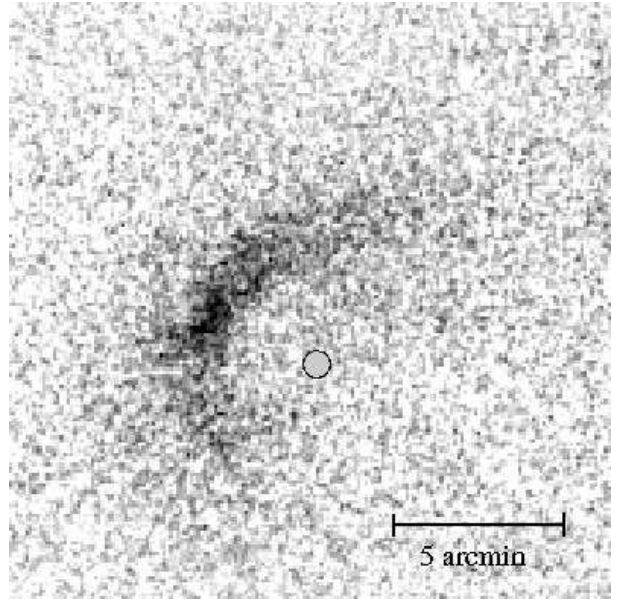}
      \caption{Image obtained in the Midcourse Space Experiment (MSX)
galactic plane survey in the $D$ medium-infrared band (13.5~$\mu$m -
15.9~$\mu$m). The position of BD$+43^\circ \ 3654$ is marked with a
grey circle. The galactic North is up and the direction of growing
galactic longitude to the left.}
         \label{msx}
\end{figure}

  The original identification of a possible bow shock associated to
BD$+43^\circ \ 3654$ was reported by van Buren \&
McCray~(\cite{vanburen88}) based on 60~$\mu$m IRAS maps, and further
details were given by van Buren et al.~(\cite{vanburen95}) and
Noriega-Crespo et al.~(\cite{noriega97}). While indeed suggestive of
a bow shock, the resolution of the IRAS~60~$\mu$m images presented
by van Buren et al.~(\cite{vanburen95}) is not sufficient to
accurately determine the shape of the bow shock and its position
with respect to the star.

  Much improved images of the region around BD$+43^\circ \ 3654$
have been provided by the Midcourse Space Experiment (MSX) satellite
(Price et al.~\cite{price01}). The BD$+43^\circ \ 3654$ bow shock
appears in them as a neat, well defined arc-shaped nebula in the $D$
(13.5-15.9~$\mu$m; see Fig.~\ref{msx}) and $E$ (18.2-25.1~$\mu$m)
bands, and is absent in the $A$ (6.8-10.8$\mu$m) and $C$
(11.1-13.2~$\mu$m) bands. The position of the apsis of the bow shock
with respect to BD$+43^\circ \ 3654$ can be well determined from
those images, being located at 3.4~arcmin from the star in a
direction that forms an angle of $62^\circ 5 \pm 10^\circ$ with the
direction towards the north galactic pole. This position angle is in
very good agreement with the position angle of the residual velocity
vector of the star (Eq.~(4)).

  The position of the bow shock with respect to the star allows us
to estimate the density of the interstellar medium through which
BD$+43^\circ \ 3654$ is moving. The apsis of the bow shock is
approximately located at the {\it stagnation radius}, which is the
distance from the star where the ram pressure of the interstellar
gas equals that of the stellar wind, given by (e.g.
Wilkin~\cite{wilkin96}):

$$R_0 = \sqrt{{{\dot M}_w v_w} \over {4 \pi \rho_a v_*^2}}  \eqno(5)$$

\noindent where ${\dot M}_w$ and $v_w$ are respectively the mass
loss rate and terminal wind velocity of the star, $\rho_a$ is the
ambient gas density, and $v_*$ is the spatial velocity of the star.
The distance given in Eq.~(5) assumes that the bow shock is bound by
shock fronts on both sides. In reality, the non-zero cooling time of
the shocked stellar wind builds up a thick layer of low-density,
high-temperature gas between the reverse shock on the stellar wind
and the bow shock. The existence of this thick layer moves the
position of the apsis of bow shock to a greater distance from the
star than that given by Eq. (5). This expression actually gives the
position of the reverse shock ahead of the star, as shown in
numerical simulations by Comer\'on \& Kaper~(\cite{comeron98b}), and
the actual position of the bow shock is normally $\sim 1.5 R_0$, the
precise distance depending on the quantities entering the right-hand
side of Eq.~(5) and the cooling curve of the stellar wind gas.
Concerning the stellar wind, we have adopted ${\dot M} =
10^{-5}$~M$_\odot$~yr$^{-1}$ and $v_* = 2300$~km~s$^{-1}$ as typical
values derived by Markova et al.~(\cite{markova04}) and Repolust et
al.~(\cite{repolust04}) for the O4I stars in their samples. Finally,
we use $v_* = 39.8$~km~s$^{-1}$ as derived in the previous Section,
assuming for simplicity that most of the velocity of BD$+43^\circ \
3654$ is on the plane of the sky and that there are no projection
effects on the position of the bow shock. Introducing these values
in Eq. (5), we obtain a number density of the local interstellar gas
$n_H \simeq 6$~cm$^{-3}$. It must be kept in mind that this is only
a rough estimate of the density, mainly due to the large
uncertainties in the values adopted for the different quantities
intervening in Eq.~(5) and the assumption that the residual motion
of the star is in the plane of the sky. In particular, we note that
Bouret et al.~(\cite{bouret05}) find mass loss rates smaller by a
factor of $\sim 3$ for the galactic O4If+ star \object{HD190429A}
when taking into account wind clumping with respect to the
homogeneous wind case, which may imply an overestimate of $n_H$ by a
similar factor due to our adopted values. In any case, the estimated
density clearly indicates that the star is moving in a tenuous
medium whose density matches well that typical of the warm HI gas in
the vicinity of the galactic midplane (e.g. Dickey \&
Lockman~\cite{dickey90}).

\section{Discussion: the origin of BD$+43^\circ \
3654$\label{origin}}

  The spectral type and estimated mass of BD$+43^\circ \ 3654$
places it among the three most massive runaway stars known to date.
The only other two comparable stars are \object{$\zeta$~Pup} and
\object{$\lambda$~Cep} (spectral types O4I(n)f and O6I(n)fp,
respectively; Ma\'\i z Apell\'aniz~\cite{maiz04}), whose masses
(65-70~M$_\odot$, as estimated by Hoogerwerf et
al.~(\cite{hoogerwerf01}) from evolutionary models by Vanbeveren et
al.~(\cite{vanbeveren98})) are similar to the one that we estimate
for BD$+43^\circ \ 3654$.

  Although currently placed near the boundary separating Cygnus~OB1
and OB9 (to the extent that this boundary may be real; see Schneider
et al.~(\cite{schneider07})), the proper motion of BD$+43^\circ \
3654$ points away from the core of Cygnus~OB2, which is
approximately marked by the location of the multiple system of O
stars \object{Cyg~OB2~8A-D}. Other early O-type stars, most notably
\object{Cyg~OB2~22A} (O3If*), \object{Cyg~OB2~22B} (O6V((f))), and
\object{Cyg~OB2~9} (O5If+), are also within few arcminutes of that
location. The position angle of BD$+43^\circ \ 3654$ with respect to
this system is $58^\circ 84$, very similar to the position angle of
its residual proper motion vector (Sect.~\ref{properties}) and of
the axis of the bow shock (Sect.~\ref{bow_shock}). In view of the
high density of very massive OB stars found in the central regions
of Cygnus~OB2 (Massey \& Thompson~\cite{massey91}), we thus consider
as a very likely possibility that BD$+43^\circ \ 3654$ was formed
there and subsequently expelled.

  Assuming that BD$+43^\circ \ 3654$ was born in the close vicinity of
Cyg~OB2~8, of which is currently separated by an angular distance
$\delta = 2^\circ 67$, the travel time to its current position is
$\tau = \delta / \sqrt{(\Delta \mu_l \cos b)^2 + (\Delta \mu_b)^2} =
1.7 \pm 0.4$~Myr, which is close to the age of the star inferred
from the evolutionary tracks and the position in the H-R diagram
(Sect.~\ref{properties}). The coincidence between the age and the
travel time supports dynamical ejection early in its life as the
cause for its runaway velocity, since there would have been no time
for a hypothetical massive companion to evolve, go through different
mass transfer episodes (Vanbeveren et al.~\cite{vanbeveren98}) and
then explode as supernova. The spatial velocities of the other two
massive runaways noted above are probably higher, unless the radial
velocity of BD$+43^\circ \ 3654$ exceeds the projected velocity on
the plane of the sky: Hoogerwerf et al.~(\cite{hoogerwerf01})
measure a velocity of 62.4~km~s$^{-1}$ for $\zeta$~Pup, and 74.0 for
$\lambda$~Cep. High spatial velocities may be the signature of an
origin by supernova ejection, since high ejection velocities by
dynamical interaction become increasingly unlikely as the mass of
the ejected star increases. The observed mass-velocity relationship
for runaway stars (Gies \& Bolton~\cite{gies86}) clearly shows this
trend. Hoogerwerf et al.~(\cite{hoogerwerf01}) favor a supernova
scenario for $\lambda$~Cep on the basis of the difference between
its age and that of the likeliest parental association. The
birthplace of $\zeta$~Pup is more uncertain according to Hoogerwerf
et al.~(\cite{hoogerwerf01}), but van Rensbergen et
al.~(\cite{vanrensbergen96}) also favor a supernova scenario. We
note however that the velocities of all three stars is well below
the upper limit for the ejection of very massive stars in encounters
with massive binaries (Leonard~\cite{leonard91}). We thus consider
the similarity between the estimated age of BD$+43^\circ \ 3654$,
and its kinematic age if it was born near the center of Cygnus~OB2,
as the strongest argument in support of a dynamical ejection,
possibly from an original cluster containing in addition Cyg~OB2~8,
9, and 22.

  BD$+43^\circ \ 3654$ is the first runaway star from Cygnus~OB2
identified thus far, but most likely it is not the only one in such
a rich association. If the fraction of runaways among O-type stars
is the same for Cygnus~OB2 as for the more nearby population of O
stars, we estimate that about ten more Cygnus~OB2 runaways may
remain to be discovered, having the potential of providing new
information on their formation environments and on the mechanisms
leading to the runaway ejection.

\begin{acknowledgements}
  It is as always a pleasure to acknowledge the support of the staff
of the Calar Alto observatory during the execution of our
observations. We also thank the detailed and constructive comments
of the referee, Dr. Dave van Buren. This research has made use of
the SIMBAD database operated at CDS, Strasbourg, France. It also
makes use of data products from the Two Micron All Sky Survey, which
is a joint project of the University of Massachusetts and the
Infrared Processing and Analysis Center/California Institute of
Technology, funded by the National Aeronautics and Space
Administration and the National Science Foundation, as well as of
data products from the Midcourse Space Experiment (MSX).  Processing
of the MSX data was funded by the Ballistic Missile Defense
Organization with additional support from NASA Office of Space
Science.  This research has also made use of the NASA/ IPAC Infrared
Science Archive, which is operated by the Jet Propulsion Laboratory,
California Institute of Technology, under contract with the National
Aeronautics and Space Administration.

\end{acknowledgements}

\end{document}